\documentclass[11pt,a4paper]{article}
\usepackage{jheppub}

\usepackage{slashed}
\usepackage{hyperref}
\usepackage{epsfig,amsmath,amsfonts,amssymb,amsthm}
\usepackage[normalem]{ulem}
\usepackage{color}
\usepackage{array}
\usepackage{multirow}
\usepackage{subfig}
\usepackage{wrapfig}
\usepackage{graphicx}
\usepackage{tabu}

\usepackage{slashed}
\usepackage{float}

\usepackage{booktabs}
\usepackage{comment}
\usepackage{enumitem}
\usepackage{graphicx}
\usepackage{siunitx}
\usepackage{xcolor}
\usepackage{verbatim}
\usepackage{dsfont}
\setlist[itemize]{leftmargin=*}

\usepackage{comment}
\usepackage{tikz}
\usetikzlibrary{decorations.pathmorphing,decorations.markings}

\tikzset{
  psi/.style={
    decoration={
      markings,
      mark=at position 0.6 with {\arrow{>}}
    },
    postaction={decorate},
    double,
    double distance=1pt
  },
  psiNoArrow/.style={
    decoration={
      markings,
      mark=at position 0.6 with 
    },
    postaction={decorate},
    double,
    double distance=1pt
  },
  nucleon/.style={
    decoration={
      markings,
      mark=at position 0.6 with {\arrow{>}}
    },
    postaction={decorate}
  },
  external/.style={},  
  gluon/.style={
  decorate, draw=black, 
  decoration={coil,amplitude=4pt, segment length=5pt}
  },
  particle/.style={draw=black, postaction={decorate}, decoration={markings,mark=at position .5 with {\arrow[draw=black]{>}}}},
 photon/.style={decorate, decoration={snake,amplitude=2pt, segment length=5pt}, draw=black}
}

\newcommand{\fftonunu}{
\begin{tikzpicture}[thick,scale=1.0]
    \draw (-1,1) -- (0,0);
    \draw (-1,-1) -- (0,0);
    \draw[dashed] (0,0) -- (2,0);
    \draw[particle] (2,0) -- (3,1);
    \draw[particle] (3,-1) -- (2,0);

    \fill[black] (0,0) circle (0.06cm);
    \fill[black] (2,0) circle (0.06cm);

    \node at (-1.25, 1.2) [external]{$\nu$};
    \node at (-1.2, -1.2) [external]{$\nu$};
    \node at (1, 0.3) [external]{$h_1,h_2$};
    \node at (3.2, 1.2) [external]{$f$};
    \node at (3.2,-1.2) [external] {$f$};
\end{tikzpicture}
}

\newcommand{\hOnehOneSchannel}{
\begin{tikzpicture}[thick,scale=1.0]
    \draw[dashed] (-1,1) -- (0,0);
    \draw[dashed] (-1,-1) -- (0,0);
    \draw[dashed] (0,0) -- (2,0);
    \draw (2,0) -- (3,1);
    \draw (2,0) -- (3,-1);

    \fill[black] (0,0) circle (0.06cm);
    \fill[black] (2,0) circle (0.06cm);

    \node at (-1.25, 1.2) [external]{$h_1$};
    \node at (-1.2, -1.2) [external]{$h_1$};
    \node at (1, 0.3) [external]{$h_1,h_2$};
    \node at (3.2, 1.2) [external]{$\nu$};
    \node at (3.2,-1.2) [external] {$\nu$};
\end{tikzpicture}
}

\newcommand{\hOnehOneTchannel}{%
\begin{tikzpicture}[thick,scale=1.0]
    \draw[dashed] (-1,0.5) -- (0,0);
    \draw (0,0) -- (1,0.5);
    \draw (0,0) -- (0,-1);
    \draw[dashed] (-1,-1.5) -- (0,-1);
    \draw (0,-1) -- (1,-1.5);

    \fill[black] (0,0) circle (0.06cm);
    \fill[black] (0,-1) circle (0.06cm);

    \node at (-1.25, 0.7) [external]{$h_1$};
    \node at (1.2, 0.7) [external]{$\nu$};
    \node at (0.3, -0.5) [external]{$\nu$};
    \node at (-1.2, -1.7) [external]{$h_1$};
    \node at (1.2,-1.7) [external] {$\nu$};
\end{tikzpicture}
}

\newcommand{\hOnehOneUchannel}{%
\begin{tikzpicture}[thick,scale=1.0]
    \draw[dashed] (-1,0.5) -- (0,-1);
    \draw (0,0) -- (1,0.5);
    \draw (0,0) -- (0,-1);
    \draw[dashed] (-1,-1.5) -- (0,0);
    \draw (0,-1) -- (1,-1.5);

    \fill[black] (0,0) circle (0.06cm);
    \fill[black] (0,-1) circle (0.06cm);

    \node at (-1.25, 0.7) [external]{$h_1$};
    \node at (1.2, 0.7) [external]{$\nu$};
    \node at (0.3, -0.5) [external]{$\nu$};
    \node at (-1.2, -1.7) [external]{$h_1$};
    \node at (1.2,-1.7) [external] {$\nu$};
\end{tikzpicture}
}

\newcommand{\be}{\begin{equation}}
\newcommand{\ee}{\end{equation}}
\newcommand{\bea}{\begin{equation}\begin{aligned}}
\newcommand{\eea}{\end{aligned}\end{equation}}

\newcommand{\hc}{\text{h.c.}}

\newcommand{\gsim}{\lower.7ex\hbox{$\;\stackrel{\textstyle>}{\sim}\;$}}
\newcommand{\lsim}{\lower.7ex\hbox{$\;\stackrel{\textstyle<}{\sim}\;$}}

\definecolor{grey}{cmyk}{0,0,0,0.75}
\definecolor{tangerine}{cmyk}{0,0.5,1,0}
\definecolor{darkgreen}{cmyk}{1,0,1,0.23}
\definecolor{Red}{rgb}{1,0,0}
\definecolor{Blue}{rgb}{0,0,1}
\definecolor{Green}{rgb}{0,1,0}
\definecolor{Grey}{cmyk}{0,0,0,0.75}
\definecolor{Tangerine}{cmyk}{0,0.5,1,0}
\definecolor{Darkgreen}{cmyk}{1,0,1,0.23}
\definecolor{Cyan}{cmyk}{1,0,0,0}
\definecolor{Yellow}{cmyk}{0,0,1,0}
\definecolor{darkblue}{cmyk}{1,0.69,0,0.11}

\newcommand{\nicpb}{Laboratory of High Energy and Computational Physics, NICPB, R\"avala pst. 10, 10143 Tallinn, Estonia}
\newcommand{\hki}{Department of Physics and Helsinki Institute of Physics, University of Helsinki, Gustaf H{\"a}llstr{\"o}min katu 2, {FI-00014} University of Helsinki, Finland}

\begin{document}

\title{Probing sterile neutrino freeze-in at stronger coupling}

\author[a]{Niko Koivunen,}
\author[b]{Oleg Lebedev,}
\author[a]{Martti Raidal}
\affiliation[a]{\nicpb} 
\affiliation[b]{\hki} 

\emailAdd{niko.koivunen@kbfi.ee}

\abstract{The regime of dark matter (DM) freeze-in at stronger coupling interpolates between   freeze-in and freeze-out. It relies on  Boltzmann-suppressed dark matter production, 
implying that the Standard Model bath temperature never exceeds the dark matter mass. In this work, we study this regime in the context of sterile neutrino dark matter, which 
can be sufficiently long-lived for a tiny sterile-active mixing.
The sterile neutrino is assumed to couple to a real singlet scalar, providing for a thermal production mechanism of the former. We find that DM mass can range from GeV to tens of TeV consistently with
all the constraints. The most interesting aspect of the consequent freeze-in phenomenology is that the sterile neutrino dark matter can be probed efficiently by both direct detection experiments
and invisible Higgs decay at the LHC.
}

\maketitle


\section{Introduction}

Understanding the nature of dark matter (DM) remains one of the main challenges of modern physics. 
If it is due to a new particle, the latter has to interact sufficiently weakly with matter and be stable or almost stable.
The sterile neutrino is an attractive dark matter candidate   \cite{Dodelson:1993je,Shi:1998km}   as long as its mixing with the active neutrinos is tiny, making it  long-lived on the cosmological scales \cite{Boyarsky:2018tvu}. The corresponding constraint as a function of its mass can be found in Ref.\,\cite{Lebedev:2023uzp}. 
In this work, we call a sterile   (or right-handed)     neutrino a fermion with the corresponding quantum numbers, i.e. neutral under the Standard Model (SM) gauge symmetries.  
 It is not necessarily responsible for the active neutrino masses   since this can be accomplished by heavier, unstable right-handed neutrinos \cite{Asaka:2005an,Asaka:2006nq}.
 While the lightest sterile neutrino has tiny couplings to leptons, it can have a significant coupling to an SM singlet $S$  \cite{Kusenko:2006rh} . The latter, in general, mixes with the Higgs field thereby providing
 us with a channel to probe dark matter via its Higgs couplings. Depending on the coupling strength and the mass,  the sterile neutrino 
 may     or may not thermalize \cite{DeRomeri:2020wng,Coy:2022unt}. If it does, its eventual abundance is determined by the standard freeze-out mechanism, possibly in the relativistic regime \cite{Lebedev:2023uzp}. In this  work, however, we focus on the non-thermal option.

 A popular alternative to the freeze-out dark matter paradigm is provided by the freeze-in mechanism \cite{Hall:2009bx}, in which case dark matter abundance builds over time without reaching its equilibrium value.
 Since DM is non-thermal, this mechanism is sensitive to its initial abundance. Although it is common to assume a zero initial value, this assumption is 
  hardly justified in standard cosmological settings due to omnipresent gravitational particle production \cite{Lebedev:2022cic}.
  In fact, the latter may be responsible for all of the {\it cold} dark matter in the form of sterile neutrinos \cite{Koutroulis:2023fgp}. Their correct abundance can be produced, for example,  by the quantum-gravity generated Planck-suppressed 
  operator which couples the sterile neutrinos $\nu$ to the inflaton $\phi$,
  \begin{equation}
  {1\over M_{\rm Pl} } \, \bar \nu \nu\, \phi^2 \;.
  \end{equation}
  Hence, the gravitational effects  are significant and require careful consideration. 
  
    The problem of gravitationally produced relics can be addressed by 
      lowering the reheating temperature, which allows for dilution of the former during the inflaton-dominated expansion era \cite{Lebedev:2022cic}.
      This   brings in
  an interesting possibility  that dark matter could be heavier than the SM bath temperature. Its production is then Boltzmann--suppressed such that freeze-in  requires larger couplings compared to those in the traditional high-$T$ scenario \cite{Cosme:2023xpa}.
  As a result, freeze-in dark matter becomes {\it observable} in direct detection experiments as well as at the LHC.\footnote{Certain non-trivial high-$T$ freeze-in models, for example, with MeV mediators and millicharged DM, can also have observable signatures \cite{Hambye:2018dpi}.} An analogous Boltzmann--suppressed regime in the context of leptogenesis was studied in \cite{Giudice:2003jh}.
  
 In this work, we study the freeze-in mechanism at stronger  coupling in the context of sterile neutrino dark matter. Its production by the SM thermal bath is due to the DM coupling to a scalar singlet mixing with the Higgs and is Boltzmann--suppressed. For a broad range of masses above 1 GeV, the sterile neutrino can account for all of the dark matter in the Universe.
   We consider separately the light and heavy DM regimes, which exhibit different 
 observational signatures, and find exciting prospects for probing freeze-in at stronger coupling.

\section{The model}
We consider an extension of 
 the SM with a real scalar singlet $S$ and a right-handed Majorana sterile neutrino $\nu$. Although one expects more than one neutrino species to be present, we focus on the lightest one in this work.
The relevant part of the neutrino Lagrangian is
\be
\mathcal{L}\supset
\frac{1}{2} M~ \overline{\nu} \nu
+\frac{1}{2}\lambda ~S~ \overline{\nu} \nu
+  \left( y_{\nu,i}\bar{L}_{L,i} \widetilde{H}\nu+\hc\right) \;,
\label{L}
\ee
where $\widetilde{H}$ is the charge-conjugate Higgs doublet. We treat the Majorana mass $M$ and the  coupling $\lambda$ as free parameters, while 
assuming 
  $y_{\nu,i}$ to be negligibly small and irrelevant to our study.\footnote{This can be justified by an approximate $Z_2$-parity symmetry $\nu \rightarrow -\nu$.}

The scalar potential is taken to be
\be
V = \mu_h^2 H^\dagger H + \frac{1}{2}\mu_s^2 S^2 
+\lambda_h (H^\dagger H)^2 + \frac{\lambda_s}{4} S^4 + \frac{\lambda_{hs}}{2} S^2 H^\dagger H \;.
\ee
Although one, in general, expects the terms   $S H^\dagger H$ and $S^3$   to be present, they are not essential for our purposes and we take them to be small.\footnote{They are generated radiatively via the $\lambda$-interaction, but can be chosen small at a given scale.}   
We consider the regime where both the
  Higgs and $S$ develop a non-zero VEV, $v_h$ and $v_s$, respectively.    
  As a result, these states mix and the mass eigenstates   $h_1, h_2$   are given by 
  \be
\left(\begin{array}{c}
h\\
S
\end{array}\right)
=
\left(\begin{array}{cc}
\cos\theta & \sin\theta\\
-\sin\theta & \cos\theta
\end{array}\right)
\left(\begin{array}{c}
h_1\\
h_2
\end{array}\right),
\ee
where $\theta$ is the mixing angle,
\be
\tan 2\theta =\frac{v_h v_s \, \lambda_{hs}}{ \lambda_s v_s^2 - \lambda_h v_h^2 }.
\ee
The 
state  $h_1$ is identified with the  $125$ GeV Higgs-like scalar observed at the LHC. In what follows,  we treat  the mass of the second scalar 
  $m_2$ along with   $\theta$ and  $v_s$ as input parameters, and present our results in terms of these. The conversion back to the Lagrangian parameters can be implemented 
  with the formulas in \cite{Falkowski:2015iwa}. Further discussion of the set-up and related phenomenology can be found in  \cite{Lebedev:2021xey}.

 The sterile neutrino $\nu$ plays the role of dark matter. As long as the active-sterile neutrino mixing is tiny, $\nu$ is long-lived on cosmological scales (see Fig.\,2 of \cite{Lebedev:2023uzp}).
 If its couplings are small, it never thermalizes and can be produced via the freeze-in mechanism. In particular, the scalars $h_1$ and $h_2$ remain in thermal equilibrium down to very low temperatures,
 and source DM production via their couplings to $\nu$. The size of the necessary couplings depends on the temperature of the SM bath $T$. In what follows, we consider the regime $M \gg T$, 
 in which case the required $\lambda$ can be as large as ${\cal O}(1)$.

\section{Boltzmann-suppressed   dark matter  production}

The reheating temperature $T_R$ of the Universe is bounded from below by about 4 MeV \cite{Hannestad:2004px} and there is no observational evidence that it has ever been high.
In fact, dilution of gravitationally produced stable relics favors low reheating temperatures \cite{Lebedev:2022cic}. This is particularly important for freeze-in models, which assume zero initial abundance 
of dark matter. This assumption is generally violated due to efficient  gravitational particle production during and immediately after inflation. 
The produced particles can however be diluted if the Universe energy density is dominated by a non-relativistic inflaton for a long time, implying a low $T_R$. In this case, a consistent
computation of freeze-in DM production becomes possible \cite{Cosme:2023xpa,Cosme:2024ndc}.

We work under the assumption that the reheating temperature of the Universe does not exceed the electroweak scale. It is then possible that the DM mass scale lies above $T_R$, in which case 
dark matter production is Boltzmann-suppressed \cite{Cosme:2023xpa}. More precisely, we assume that $M$ is above the SM bath temperature at any stage of the Universe evolution, i.e. above  the maximal temperature
(see the discussion in \cite{Cosme:2024ndc}).
Dark matter is  produced entirely via thermal emission and the exact relic abundance depends on the profile  of the temperature evolution.
A simple benchmark approximation is obtained by assuming {\it instant reheating}, i.e. that the SM bath temperature increases abruptly from 0 to $T_R$. 
Although this situation is idealized, it gives a good approximation to classes of models with flat temperature profiles before reheating. By rescaling the results, it can be applied to more general situations,
as   detailed in  \cite{Cosme:2024ndc}.
 
 The production modes and signatures depend on the DM mass $M$. For heavy dark matter, it is produced primarily via scattering of the Higgs-like scalars $h_1$ and $h_2$. Light dark matter, on the other hand, is more efficiently produced via SM fermion annihilation. In what follows, we consider these regimes separately.\footnote{We leave out the intermediate mass regime, whose treatment requires more sophisticated numerical tools, e.g. an upgraded version of  micrOMEGAs \cite{Belanger:2013ywg}.}

 \subsection{Heavy dark matter}
 
 This regime is defined by $M^2 \gg m_h^2$. In practise, we take $M>250\,$GeV.  Since the bath temperature is far below $M$, only the most energetic particles from the Boltzmann tail can cause 
 DM production. The main production modes are $h_i h_j \rightarrow \nu\nu$, while the gauge boson fusion is subleading. Unlike Refs.\,\cite{Drewes:2015eoa,Konig:2016dzg}, we consider heavy DM such that $h_{1,2} $ decays cannot produce it
 for kinematic reasons.
  
 Both $h_1$ and $h_2$ remain in thermal equilibrium down to very low temperatures. This is due to the efficient reaction $\bar f f \leftrightarrow h_{1,2}$, where 
 $f$ is a light SM fermion. It is faster than the expansion of the Universe 
 at temperature $T$ 
 as long as 
 \begin{equation}
y_{fi} \sqrt{m_{h_i} M_{\rm Pl}}  \gtrsim T \;,
\label{therm}
\end{equation}
omitting unessential factors (below 10). Here $y_{fi}$   ($i=1,2$) is the corresponding scalar-fermion Yukawa coupling, $\sqrt{2} \sin\theta    \,m_f/v_h$ or $\sqrt{2} \cos\theta    \,m_f/v_h$. 
This implies that thermal equilibrium is maintained if
   ${\cal O}\left( 10^{10} \right) \, y_{fi} > T /\,$GeV, 
   such that for all     parameter choices of interest    ($\theta \sim 10^{-1}, m_2 < \,$TeV)  both Higgs-like scalars  are thermalized. 
 
 \subsubsection{The Boltzmann equation}
 
 The DM number density is determined by the Boltzmann equation. We would like to write it in a convenient form applicable to both pure freeze-in regime, where the sterile neutrinos
 may not be in kinetic equilibrium with the SM bath, and the regime where DM annihilation becomes tangible and kinetic equilibrium is assumed \cite{Kolb:1990vq}.

 Focussing on the Higgs contributions, we have 
\begin{eqnarray}
\dot n_\nu +3Hn_\nu & = & 2 \,\Gamma (h_i h_j \rightarrow \nu\nu) - 2\,\Gamma (\nu \nu \rightarrow h_i h_j)  \\
& =&  
2\, \Bigl(     \Gamma (h_1 h_1 \rightarrow \nu \nu) +\Gamma (h_1 h_2 \rightarrow \nu \nu) +\Gamma (h_2 h_2 \rightarrow \nu \nu) \Bigr)
\; \left( 1-n_\nu^2/n_{\nu}^{\rm eq\,2}     \right)\, ,
\nonumber 
\end{eqnarray}
where we have used the Boltzmann distribution for computing the rates\footnote{The quantum statistics effects  are very small for particles at the Boltzmann tail.}
and the second equation assumes kinetic equilibrium.
$\Gamma$ denotes the reaction rate per unit volume. 
The factor of 2 appears due to particle number change by 2 units in each reaction.
In the pure freeze-in regime, the annihilation rate is negligible, i.e. the term proportional to $n^2_\nu$ can be dropped. As the coupling grows, 
the $h_i-\nu$ system reaches kinetic equilibrium, in which case the production and annihilation rates 
are related via the chemical potential. At yet larger couplings, the system thermalizes fully and the right hand side vanishes, until DM freeze-out.
  
 The above Boltzmann equation can be put in a more conventional form using  \cite{Gondolo:1990dk}
  \begin{equation}
\Gamma(ab \rightarrow cd)= (2\pi)^{-6} \int d^3p_a d^3 p_b\, f(p_a) f(p_b) \; \sigma (p_a,p_b) \,v_{M\o{}l}  =  n_a n_b \langle \sigma \, v_{M\o{}l} \rangle \;,
\label{G}
\end{equation}
where  $f$ is the momentum distribution function,
$\sigma$ is the reaction cross section which includes the phase space symmetry factors for $both$ initial and final particles, $v_{M\o{}l} $ is the M\o{}ller velocity and $\langle ... \rangle$ denotes thermal averaging.  
In kinetic equilibrium,
$   f = e^{-{E-\mu \over T}}$ with $\mu$ being the chemical potential \cite{Kolb:1990vq,Bernstein:1985th}. This, together with energy conservation, implies
  \begin{equation}
\Gamma (h_i h_j \rightarrow \nu \nu)=       e^{-2\mu/T}\, \Gamma   (\nu\nu \rightarrow  h_i h_j )   ~~,~~ n_\nu/n_{\nu}^{\rm eq}= e^{\mu/T} \;.
\end{equation}
 Then, the Boltzmann equation can be written, for example, in terms of neutrino annihilation cross sections, 
  \begin{eqnarray}
\dot n_\nu + 3Hn_\nu &=& 2\, \Bigl(     \langle \sigma   ( \nu\nu \rightarrow h_1 h_1)     v \rangle       +    \langle \sigma   ( \nu\nu \rightarrow h_1 h_2)     v \rangle   +  \langle \sigma   ( \nu\nu \rightarrow h_2 h_2)     v \rangle \Bigr) \, \nonumber\\
&\times& \left(n_{\nu}^{\rm eq\,2}-n_\nu^2 \right) 
\;.
\end{eqnarray}
 The term proportional to  $n_{\nu}^{\rm eq\,2}$ is responsible for DM production and  valid even in the pure freeze-in regime at feeble couplings. Despite a somewhat deceiving appearance, it {\it does not} assume that sterile neutrinos are in any kind of equilibrium with the 
 environment. 
 Although the thermal averaging prescription formally requires using the Boltzmann distribution for $\nu$, the full phase space integration makes sure that 
  the resulting $ \langle \sigma   ( \nu\nu \rightarrow h_i h_j)     v \rangle \, n_{\nu}^{\rm eq\,2}$ is just the rate of $h_i h_j \rightarrow \nu \nu$ with thermal $h_i , h_j$.
 In contrast, the annihilation term proportional to $n_\nu^2$ assumes chemical potential factorization and, thus, kinetic equilibrium for the $\nu-h_i$ system.
 This is justified since kinetic equilibrium sets in before DM annihilation becomes important, due to $n_{h_i} \gg n_\nu$.

  \subsubsection{Relic abundance of dark matter and constraints}

  \begin{figure}[t]
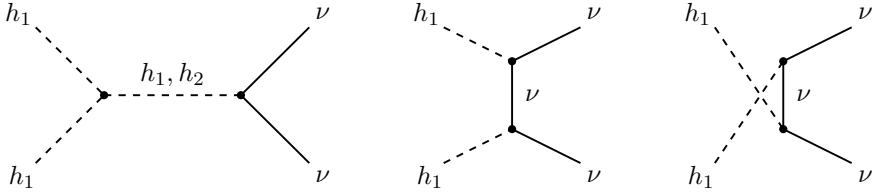

    \centering
    \scalebox{0.9}{\hOnehOneSchannel 
    \hspace{0.7cm}
    \hOnehOneTchannel
    \hspace{0.7cm}
    \hOnehOneUchannel}
    \caption{Sterile neutrino production via scalar annihilation $h_1 h_1\to \nu\nu$. (Analogous diagrams exist for the $h_2 h_2\to \nu\nu$ and $h_1 h_2\to \nu\nu$ channels.)}
    \label{production diagrams}
\vspace*{-2mm}
\end{figure}

  The sterile neutrino production processes are shown in Fig.\,\ref{production diagrams}. At small $\lambda$, the leading contribution comes from the diagrams involving a triple scalar vertex. It is instructive to consider the limit $\theta \ll 1$, in which case the $h_2$ annihilation via $\Delta {\cal L}=\lambda_{222} \,h_2^3$ dominates and 
  \begin{eqnarray}
  \sigma_{h_2 h_2 \rightarrow \nu \nu} = {1\over 2} \times {9\,\lambda^2 \lambda_{222}^2 \over 4 \pi s^3}\, {   (s-4 M^2)^{3/2} \over (s-4 m_2^2)^{1/2}  } \;.
  \end{eqnarray}   
  Here we factor out the  1/2 due to identical particles in the initial state, which is included directly in $\sigma$ in our convention. $s$ is the Mandelstam variable and $\lambda_{222} \rightarrow 
  \lambda_s v_s$ in the small mixing limit. We observe the velocity suppression $v^3$ of the result, as expected for production of non-relativistic fermions. The above expression agrees with 
  the result  of \cite{DeRomeri:2020wng}.
  
  The Boltzmann equation can be solved analytically  at $M \gg T$    with the above approximations.
  Using the Gelmini-Gondolo result \cite{Gondolo:1990dk}, the rate (\ref{G}) takes the form
  \begin{equation}
  \Gamma(h_2 h_2 \rightarrow \nu \nu) \simeq {27\, \lambda^2 \lambda_{222}^2 \, T^4 \over 2^{10} \pi^4 M^2    } \, e^{-2M/T} \;.
  \end{equation}
  Here we have used the limit $2M \gg T$ and the corresponding expansion of the Bessel function.
  Particle production takes place for a short time close to the maximal temperature $T_R$. Therefore, we may take the number of degrees of freedom $g_*$ approximately constant in this period
  and the Boltzmann equation in the $pure$ freeze-in regime takes the form
  \begin{equation}
  T^3 \, {d\over dt}  \left( {n_\nu \over T^3}   \right)=2 \, \Gamma(h_2 h_2 \rightarrow \nu \nu) \;.
  \end{equation} 
  The equation can easily be solved analytically and the resulting DM relic abundance reads
  \begin{equation}
  Y \equiv {n_\nu \over s_{\rm SM}} \simeq 1.7 \times 10^{-6} \; {\lambda^2 \lambda_{222}^2 \, M_{\rm Pl} \over M^3} \, e^{-2M/T_R}\;,
  \end{equation}
  where $s_{\rm SM}$ is the Standard Model entropy density, $s_{\rm SM}= {2\pi^2 \over 45} g_* T^3$ and $g_* \simeq 107$.
  The observational constraint $Y= 4.4 \times 10^{-10}\, {\rm GeV}/M$ then requires
  \begin{equation}
  {\lambda \,\lambda_{222} \over M}\; e^{-M/T_R} \simeq 10^{-11} \;.
  \label{appr}
  \end{equation}
  Therefore, $M$ depends logarithmically on the coupling. Since  $M \gg T_R$, 
  order one $\lambda$ is consistent with freeze-in.
  For typical  
  $\lambda_{222} \sim \lambda_s v_s$ of order 100 GeV, the neutrino coupling is constrained by ${\lambda  \over M/{\rm GeV}}\; e^{-M/T_R} \sim 10^{-13}$.
 Thus, for $\lambda\sim 1$ and $M\sim 100\,$GeV, the reheating temperature is about $M/25$.

      \begin{figure}[h!]
    \centering
    \includegraphics[width=0.49\textwidth]{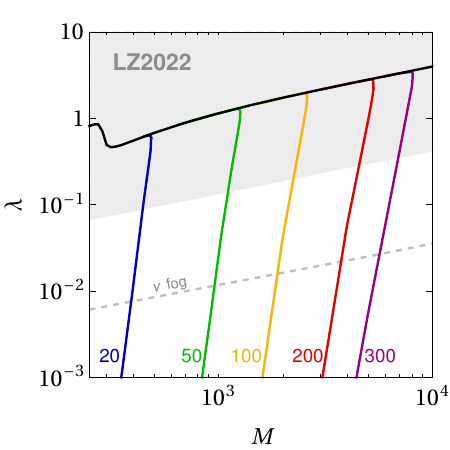}
    \includegraphics[width=0.465\textwidth]{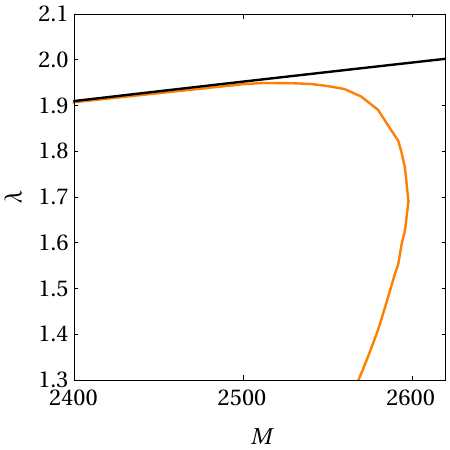}
    \caption{Parameter space for heavy  sterile neutrino dark matter with $\sin\theta=0.2, m_2 = 300 \,{\rm GeV}, v_s=200 \,{\rm GeV}$ ($|\lambda_{222}|=213\,$GeV).  {\it Left:} 
    the colored curves correspond to the correct relic abundance for a given $T_R$ in GeV. The black line represents thermal dark matter freeze-out. The shaded area is ruled out by direct DM detection (LZ2022), while the dashed line shows the neutrino background.      {\it Right:} zoomed-in transition region  from freeze-in to freeze-out for $T_R=100\,$GeV (yellow line).}
    \label{par-space}
\end{figure}

   As the coupling increases, the DM annihilation term becomes important. It reduces the DM density compared to the pure freeze-in approximation and eventually leads to thermalization. In the latter case,
   $\Gamma (h_i h_j \rightarrow \nu\nu) = \Gamma (\nu \nu \rightarrow h_i h_j)$ until freeze-out, which determines the DM relic abundance. Our full numerical results are presented in Fig.\,\ref{par-space}.
  We find that, in the parameter range of interest, the $\lambda_{222}$ contribution considered above  indeed dominates, even at larger $\lambda$.
    The colored lines in the figure correspond to the correct DM relic abundance for a fixed $T_R$. At sufficiently large coupling, they all merge with the thermal freeze-out curve (black). The latter exhibits a kink at $M=m_2$
    due to the annihilation channel $\nu\nu \rightarrow h_2 h_2$ becoming available. 
    
    The transition region  from pure freeze-in to freeze-out is shown in the right panel of Fig.\,\ref{par-space}. We see that increasing the coupling beyond a certain limit does not increase the abundance anymore,  thermalization sets in and the dependence on $T_R$ is lost. The transition is smooth and shows how the fundamentally different DM regimes are continuously connected, as also observed in Ref.\,\cite{Silva-Malpartida:2023yks}.

   The most important constraint on the model is imposed by the direct DM detection bounds \cite{LZ:2022lsv}.
   The DM-nucleon scattering cross section is given by
  \be
\sigma_{\rm SI} = \frac{  \lambda^2   \mu_N^2}{\pi}\left(\frac{f_N m_N}{v_h}\right)^2 \left(\frac{1}{m_1^2}-\frac{1}{m_2^2}\right)^2 \sin^2\theta\cos^2\theta  \, ,
\label{DD}
\ee
where $m_N$ is the nucleon mass, $\mu_N \simeq 1\,$GeV is the $\nu$-$m_N$ reduced mass and $f_N\approx 0.3$.

   In fermionic dark matter models of the Higgs portal type \cite{Kim:2006af,Kim:2008pp,Baek:2011aa}, the direct detection constraint   (along with the LHC bound)   is so strong that it essentially rules out the thermal DM option \cite{Djouadi:2011aa}, unless CP violation is allowed \cite{Lopez-Honorez:2012tov}. This is mainly due to velocity suppression of DM annihilation, while $\sigma_{\rm SI}$ is not subject to such suppression.
  In the case of Boltzmann-suppressed freeze-in, on the other hand,  all the constraints are satisfied for couplings below $10^{-1}-1$ in the mass range ${\cal O}(10^2)\,$GeV--10 TeV.  This parameter space will be continuously probed by
  current and future direct detection experiments such as XENONnT \cite{XENON:2020kmp} and DARWIN\cite{DARWIN:2016hyl}. Beyond 10 TeV, the sensitivity of these experiments drops, however freeze-in at stronger coupling allows for much higher masses without conflicting perturbativity. Another limitation for the above searches is given by the neutrino background scattering marked by ``$\nu$ fog'' in the figure.

   We have chosen $\sin\theta=0.2$ and $m_2=300\,$GeV as a benchmark parameter choice, motivated by the LHC constraints. The LHC searches for a ``heavy Higgs'' continuously set stricter bounds
   on the Higgs-like resonances (see e.g. \cite{ATLAS:2022eap}). A recent combination of the LHC constraints on $\theta, m_2$ can be found in 
   \cite{Robens:2022cun,Tania2} and our benchmark represents a reasonable choice.

   The scaling of the results with $\theta$ and $m_2$ is readily seen from Eqs.\,\ref{appr},\,\ref{DD}. The relic abundance is independent of $\theta$ and $m_2$ as long as $h_2$ remains in thermal equilibrium,
   while it is sensitive to the trilinear coupling $\lambda_{222}$. The direct detection constraint is primarily due to the exchange of $h_1$, so it is effectively independent of $m_2$. On the other hand,
   it relaxes  quickly as $\theta$ decreases: the minimal excluded coupling  scales as  $\lambda \propto 1/\theta$.  
   Therefore, our results for other values of $\theta$ and $m_2$ can be deduced from Fig.\,\ref{par-space}.

  In the above analysis, we have focussed on the DM production via scalar annihilation.  
The  SM fermion annihilation  channel  is suppressed by the Yukawa couplings and can be neglected. Indeed, although there are many more fermions than the Higgses in the thermal bath,
  only few of them   ($\propto e^{-M/T}$)    have enough energy to produce DM. This gives no advantage  over the Higgs channel, while there is the Yukawa suppression factor.

\subsection{Light dark matter}

\begin{figure}[t]
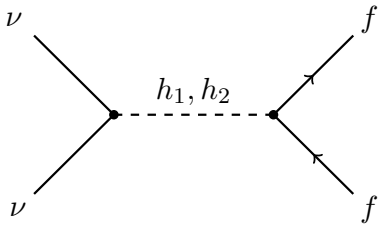

    \centering
    \scalebox{1.05}{\fftonunu 
    }
    \caption{ The leading channel for light dark matter production/annihilation  ($f=b, c,\tau$).}
    \label{low mass production diagrams}
\vspace*{-2mm}
\end{figure}

This regime corresponds to $M \ll m_h$. Since the Boltzmann suppression for DM production is $e^{-M/T}$, it is milder than suppression of the Higgs abundance in the thermal bath $e^{-m_h/T}$ and the Higgs scattering or decay cannot be the leading production channel. The main process is instead $\bar f f \rightarrow \nu\nu$, where $f$ is a light SM fermion (Fig.\,\ref{low mass production diagrams}).

Analogously to the heavy DM case, the Boltzmann equation can be put in the form
\bea
\dot n_\nu +3Hn_\nu & = 2\,
\Big(
 \langle \sigma_{\nu\nu\to b\bar b} v\rangle
+\langle \sigma_{\nu\nu\to c\bar c} v\rangle
+\langle \sigma_{\nu\nu\to \tau\bar\tau} v\rangle\Big)\Big[(n_\nu^{eq})^2 -n_\nu^2\Big].
\eea
Here the spin-averaged cross section is
\be
\sigma_{\nu\nu\to f\bar f} = {1\over 2}\times \frac{N_c\, y_f^2 \,\lambda^2}{32\pi s}(s-4m_f^2)^{3/2}(s-4M^2)^{1/2}\left[\frac{1}{s-m_1^2}-\frac{1}{s-m_2^2}\right]^2
\sin^2\theta \, \cos^2\theta,
\ee
with $f=b,c,\tau$ and $N_c=3$ for $b$ and $c$, and $1$ for $\tau$. The Yukawa coupling $y_f$ is defined analogously to that in Eq.\,\ref{L} and 
the factor of $1/2$ is due to identical particles in the initial state, which we include directly in $\sigma$. Note that $n_{\nu}^{eq}$ includes 2 spin degrees of freedom.

The analysis proceeds largely parallel to the heavy DM case since the temperature dependence of the reaction rate remains the same, $T^4\, e^{-2M/T}$. 
We take the lowest $T_R$ to be 200 MeV so that the thermal production calculation in terms of elementary fermions is adequate. Below this temperature,
the QCD phase transition effects become strong. 

The resulting abundance for  $m_2^2 \gg m_1^2$ and neglecting the SM fermion mass is
\begin{equation}
Y \simeq \times 10^{-5} \, N_c\,  y_f^2 \lambda^2 \sin^2 \theta \, \cos^2 \theta \, {M^3 M_{\rm Pl} \over m_1^4} \, e^{-2M/T_R} \, .
\end{equation}
Here we have used $g_* \simeq 70$ for temperatures around the GeV, while  in general the abundance scales as $g_*^{-3/2}$. 
When the bottom quark dominates, this sets the constraint
\begin{equation}
\lambda \, (M/ {\rm GeV})^2 \, e^{-M/T_R} \simeq 1.5 \times 10^{-6}  / \sin\theta\;. 
\end{equation}
In reality, the fermion masses are not negligible  and the RHS of this equation receives the correction factor 
$(1- (m_f/M)^2)^{-3/4}$.  
Order 1 coupling then requires the hierarchy between $M$ and $T_R$ to be somewhat milder than that  in the heavy DM case: for typical values $M/T_R \sim 15$.

 \begin{figure}[h!]
    \centering
    \includegraphics[width=0.55\textwidth]{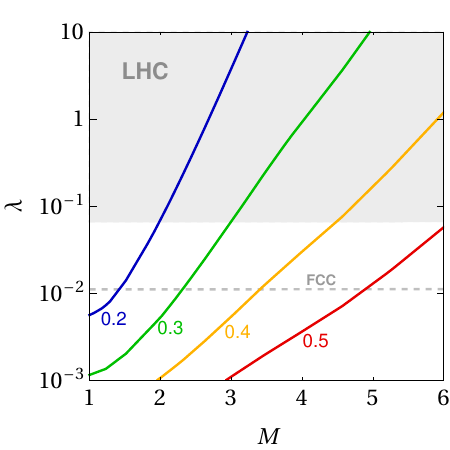}
      \caption{Parameter space for light  freeze-in  DM at 
      $\sin\theta=0.2$.    The colored curves correspond to the correct relic abundance for a given $T_R$ in GeV. The shaded area is excluded by the LHC measurement of the Higgs invisible decay,
      while the dashed line shows the FCC prospects in this channel.}
    \label{par-space1}
\end{figure}

For $M< 6\,$GeV, the direct DM detection constraints become very loose and the main bound is imposed by the invisible Higgs decay $h_1\to \nu\nu$ \cite{Djouadi:2011aa}.
The corresponding branching ratio is given by 
\be
\rm{BR}_{\rm inv}=\frac{\Gamma_{h_1\to \nu\nu}}{\cos^2\theta \, \Gamma^{tot}_{SM}+\Gamma_{h_1\to \nu\nu}} \;,
\ee
where $\Gamma^{\rm tot}_{\rm SM}$ is the SM total Higgs width.
The decay width into the Majorana sterile neutrinos is 
\be
\Gamma_{h_1\to \nu\nu} = \frac{(m_1^2-4M^2)^{3/2}}{16\pi m_1^2} \sin^2\theta~\lambda^2\, .
\ee
A detailed discussion of the $h_1$ invisible decay can be found in \cite{Biekotter:2022ckj}, while for our purposes the above formulae suffice. This observable as a probe of freeze-in was also
discussed in  \cite{Bringmann:2021sth}.

Our results are presented in Fig.\,\ref{par-space1}. We restrict ourselves to the $M$ range between 1 and 6 GeV so that the direct DM detection constraint can be dropped, yet
$T_R$ is not too low to avoid  complications of the QCD phase transition. The colored curves correspond to the correct DM relic abundance due to freeze-in. The annihilation effect
is never important in our parameter range and the thermal abundance would be reached for $\lambda \gg 1$ only, as we have verified numerically. Above the bottom quark threshold, $M > m_b$,
$b$-quark annihilation dominates DM production, while for lower masses, this channel gets suppressed by $e^{-2m_b/T}$ instead of $e^{-2M/T}$. Hence, 
the $c$-quark makes the main contribution. 
Below the charm threshold ($M<1.3\,$GeV), the $s$-quark dominates and 
  our results become less reliable due to the uncertainty in the $s$-quark Yukawa coupling. In Fig.\,\ref{par-space1}, we use $m_s \sim 95\,$MeV.

 We impose the  bound of  $\rm{BR}_{inv} \lesssim 0.1$ \cite{ATLAS:2020kdi} to set the benchmark.\footnote{The exact bound depends on the treatment of the Higgs signal strength $\mu$ 
 and the ATLAS + CMS result combination.}  This excludes couplings above $10^{-1}$. The HL-LHC will improve the bound on the Higgs coupling strength and   $\rm{BR}_{inv} $
by a factor of a few \cite{RivadeneiraBracho:2022sph}, which allows one to probe the coupling region somewhat below the exclusion line, e.g. by a factor of two. The future collider FCC tuned as a Higgs factory can reach the 
limit of $\rm{BR}_{inv} \sim 0.003$ (see, e.g.\,\cite{FCC}), corresponding to the dashed line in Fig.\,\ref{par-space1}.

Our  results shown in Fig.\,\ref{par-space1} apply to other input parameter choices by a simple rescaling. First, $h_2$ plays no role at low energies. Second, the combination that enters the observables 
at $\theta^2 \ll 1$ is
$\lambda \, \sin\theta$. Hence, changing $\theta$ amounts to a rescaling $\lambda$ in this limit.

To conclude, contrary to the thermal dark matter case, there are   prospects of observing light non-thermal  DM via invisible Higgs decay at both the LHC and FCC. Improvements in its measurement
will probe smaller $\nu$ couplings in the freeze-in regime.

\subsection{Beyond the instant reheating approximation}

Our results have been obtained under  the simplifying assumption that the SM bath temperature increases from zero to $T_R$ instantaneously.
Although this does not appear as a very realistic assumption, our results apply, either directly or via appropriate rescaling,  to a broad class of models, where the SM temperature stays constant over cosmological times \cite{Cosme:2024ndc}.
In such models, the SM radiation is produced by decay of a field subdominant in the energy balance, instead of the inflaton $\phi$.  This applies, for example, to the decay cascade $\phi \rightarrow \chi \chi \rightarrow {\rm SM }$, where $\chi$ is a feebly interacting field. In this case, the temperature of the SM bath remains constant after reaching its maximum, until reheating. DM production occurs most efficiently at the maximal temperature $T_{\max  }$, due to the exponential suppression of the rate. When the reheating and maximal temperatures coincide $T_{\max  }\simeq T_R$, our results in terms of the $\left(  M, \lambda \right)$ parameters  apply    directly, with a small correction \cite{Cosme:2024ndc}. 

In general, freeze-in at stronger coupling operates as long as the maximal temperature is well below the DM mass and $T_R <T_{\max  } $. The corresponding DM relic abundance  $Y$ can be obtained by rescaling our results \cite{Cosme:2024ndc}. 
First, one replaces $T_R \rightarrow T_{\max  } $ in the exponent since DM production peaks at that point. Second, one accounts for dilution of DM (entropy production) in the period between 
$T_{\max  } $ and $T_R$. The latter depends on the scaling of the Hubble rate and $T$ in this period, hence the conclusion  is model dependent. 
 The general result is thus 
 \begin{equation}
 Y(T_R) \rightarrow Y(T_{\rm max}) \times \left(   {T_R \over T_{\rm max}}  \right)^\alpha \;, 
 \end{equation}
where $ Y(T_R) $ represents our result and $\alpha >0$ is model dependent (see \cite{Cosme:2024ndc} for details). Since the abundance scales as $\lambda^2$, the coupling gets rescaled by the square root of the above factor.
The result is therefore model-dependent, which is inherent in any non-thermal DM model. For example, the conventional  high-$T$ freeze-in models suffer from the  uncontrollable gravitational particle production background \cite{Lebedev:2022cic}.

Generally, there is no evidence in favor of  a specific mechanism of the SM particle generation or a particular range of $T_R$.  In this work, we take an agnostic stance on this matter and treat DM production from the effective low energy perspective. Our assumptions are that the initial DM abundance is negligible and that DM is primarily produced by the SM thermal bath. These may or may not be realistic in a given UV complete framework.

\section{Conclusion}

Freeze-in at stronger coupling occurs when the dark matter mass exceeds the Standard Model bath temperature. The dark matter coupling is allowed to be as large as order one, without thermalizing the system. 
At yet larger couplings, dark matter equilibrates with the SM thermal bath and one recovers the usual DM freeze-out. In this sense, the regime of freeze-in at stronger coupling interpolates between freeze-in and freeze-out.

In this work, we have studied    sterile neutrino  DM freeze-in at stronger coupling.         It couples to a singlet scalar, which mixes with the Higgs and thus allows for thermal production of sterile neutrinos.
Although the usual DM freeze-out is ruled by the direct detection constraints, freeze-in at stronger coupling is consistent with the bounds and can account for the observed (cold) DM density. 
The scenario is currently being probed by the direct detection experiments and, as their sensitivity improves down to the ``neutrino floor'', 
lower couplings and higher reheating temperatures will be continuously explored. For light (GeV scale) dark matter, the main signature is provided by the invisible Higgs decay at the LHC. Part of the parameter space is 
already excluded, while lower couplings corresponding to the Higgs ${\rm BR_{inv}}$  at the percent level and below will be probed at the HL-LHC and the FCC.

 We note that the scenario considered is well theoretically motivated. Indeed, the usual freeze-in models suffer from the problem of initial conditions: gravitational inflationary and postinflationary particle production creates a strong background for the abundance calculations. The problem can be addressed by lowering the reheating temperature, in which case the dark matter mass scale could readily  exceed the
 SM bath temperature. Then,  freeze-in at stronger coupling would be the leading mechanism for DM production.
\\ \ \\
{\bf Acknowledgements.} OL is grateful to Tania Robens for useful communications.  
This work was supported by the Estonian Research Council grants PRG803, PRG1677, RVTT3, RVTT7, and the CoE program TK202 ``Fundamental Universe'’.

\appendix

\section{Appendix}\label{extra}

In this Appendix, we list some of the analytical results for the cross sections we use in our study. 
The thermal averaged cross section for the process $h_i h_j\to \nu\nu$ reads 
\be
\langle \sigma_{h_i h_j\to \nu\nu} v\rangle
=\frac{1}{n_{h_i}^{eq}n_{h_j}^{eq}}\frac{T}{32\pi^4}
\int_{4M^2}^\infty ds \big[s-(m_i+m_j)^2\big]
\big[s-(m_i-m_j)^2\big]\frac{\sigma_{h_i h_j\to \nu\nu} }{\sqrt{s}}
K_1(\sqrt{s}/T).
\ee
Defining  the scalar trilinear  couplings as
\begin{equation}
{\cal L}_{ijk}= \lambda_{ijk} \, h_i h_j h_k \;,
\end{equation}
we find the following 
  production cross sections  (including the $initial$ state  phase space symmetry factors):
\bea
\sigma_{h_1 h_1\to \nu\nu} = \frac{1}{128\pi s}\frac{\sqrt{s-4M^2}}{\sqrt{s-4m_1^2}}\left[A_{11}+B_{11}~\log\left(\frac{s-2m_1^2+\sqrt{s-4m_1^2}\sqrt{s-4M^2}}{s-2m_1^2-\sqrt{s-4m_1^2}\sqrt{s-4M^2}}\right)\right],
\eea
where 
\bea
A_{11} & = 4 (s-4M^2)\left(\frac{(-\sin\theta\lambda) (3!~\lambda_{111})}{s-m_1^2}+\frac{(\cos\theta\lambda) (2!~\lambda_{112})}{s-m_2^2}\right)^2\\
&-32 M (-\sin\theta \lambda)^2 \left(\frac{(-\sin\theta\lambda) (3!~\lambda_{111})}{s-m_1^2}+\frac{(\cos\theta\lambda) (2!~\lambda_{112})}{s-m_2^2}\right)\\
& -\frac{8(-\sin\theta\lambda)^4(2sM^2+16M^4-16M^2 m_1^2+3m_1^4)}{s M^2 -4M^2 m_1^2 +m_1^4},
\eea
\bea
B_{11} & = -\frac{16 M (-\sin\theta\lambda)^2(s-8M^2+2m_1^2)}{\sqrt{s-4m_1^2}\sqrt{s-4M^2}}
\left(\frac{(-\sin\theta\lambda) (3!~\lambda_{111})}{s-m_1^2}+\frac{(\cos\theta\lambda) (2!~\lambda_{112})}{s-m_2^2}\right)\\
& + \frac{8(-\sin\theta\lambda)^4(s^2+16sM^2-32M^4+6m_1^4-4sm_1^2-16M^2m_1^2)}{(s-2m_1^2)\sqrt{s-4m_1^2}\sqrt{s-4M^2}},
\eea

\bea
\sigma_{h_2 h_2\to \nu\nu} = \frac{1}{128\pi s}\frac{\sqrt{s-4M^2}}{\sqrt{s-4m_2^2}}\left[A_{22}+B_{22}~\log\left(\frac{s-2m_2^2+\sqrt{s-4m_2^2}\sqrt{s-4M^2}}{s-2m_2^2-\sqrt{s-4m_2^2}\sqrt{s-4M^2}}\right)\right],
\eea
where 
\bea
A_{22} & = 4 (s-4M^2)\left(\frac{(-\sin\theta\lambda) (2!~\lambda_{221})}{s-m_1^2}+\frac{(\cos\theta\lambda) (3!~\lambda_{222})}{s-m_2^2}\right)^2\\
&-32 M (\cos\theta \lambda)^2 \left(\frac{(-\sin\theta\lambda) (2!~\lambda_{221})}{s-m_1^2}+\frac{(\cos\theta\lambda) (3!~\lambda_{222})}{s-m_2^2}\right)\\
& -\frac{8(\cos\theta\lambda)^4(2sM^2+16M^4-16M^2 m_2^2+3m_2^4)}{s M^2 -4M^2 m_2^2 +m_2^4},
\eea
\bea
B_{22} & = -\frac{16 M (\cos\theta\lambda)^2(s-8M^2+2m_2^2)}{\sqrt{s-4m_2^2}\sqrt{s-4M^2}}
\left(\frac{(-\sin\theta\lambda) (2!~\lambda_{221})}{s-m_1^2}+\frac{(\cos\theta\lambda) (3!~\lambda_{222})}{s-m_2^2}\right)\\
& + \frac{8(\cos\theta\lambda)^4(s^2+16sM^2-32M^4+6m_2^4-4sm_2^2-16M^2m_2^2)}{(s-2m_2^2)\sqrt{s-4m_2^2}\sqrt{s-4M^2}},
\eea

\bea
&\sigma_{h_1 h_2\to \nu\nu} = \frac{1}{64\pi s}\frac{\sqrt{s-4M^2}}{\sqrt{s-2(m_1^2+m_2^2)+(m_1^2-m_2^2)^2/s}}\\
&\left[A_{12}+B_{12}~\log\left(\frac{s-m_1^2-m_2^2+\sqrt{s-4M^2}\sqrt{s-2(m_1^2+m_2^2)+(m_1^2-m_2^2)^2/s}}{s-m_1^2-m_2^2-\sqrt{s-4M^2}\sqrt{s-2(m_1^2+m_2^2)+(m_1^2-m_2^2)^2/s}}\right)\right],
\eea
where 
\bea
A_{12} & = 4 (s-4M^2)\left(\frac{(-\sin\theta\lambda) (2!~\lambda_{112})}{s-m_1^2}+\frac{(\cos\theta\lambda) (2!~\lambda_{221})}{s-m_2^2}\right)^2\\
&-32 M (-\sin\theta \lambda)(\cos\theta\lambda) \left(\frac{(-\sin\theta\lambda) (2!~\lambda_{112})}{s-m_1^2}+\frac{(\cos\theta\lambda) (2!~\lambda_{221})}{s-m_2^2}\right)\\
& -\frac{8(-\sin\theta\lambda)^2(\cos\theta\lambda)^2\big[2sM^2+16M^4-8M^2 (m_1^2+m_2^2)+3m_1^2m_2^2-2M^2(m_1^2-m_2^2)/s\big]}{s M^2 -2M^2 (m_1^2+m_2^2) +m_1^2 m_2^2+M^2(m_1^2-m_2^2)^2/s},
\eea

\begin{eqnarray}
\resizebox{1.0 \textwidth}{!} 
{
$ 
\begin{array}{l}
 B_{12}  = -\frac{16 M (-\sin\theta\lambda)(\cos\theta\lambda)(s-8M^2+m_1^2+m_2^2)}{\sqrt{s-4M^2}\sqrt{s-2(m_1^2+m_2^2)+(m_1^2-m_2^2)/s}}
\left(\frac{(-\sin\theta\lambda) (2!~\lambda_{112})}{s-m_1^2}+\frac{(\cos\theta\lambda) (2!~\lambda_{221})}{s-m_2^2}\right)\\
 + \frac{8(-\sin\theta\lambda)^2(\cos\theta\lambda)^2\big[s^2+16sM^2-32M^4+m_1^4+4m_1^2m_2^2+m_2^4-2s(m_1^2+m_2^2)-8M^2(m_1^2+m_2^2)\big]}{(s-m_1^2-m_2^2)\sqrt{s-4M^2}\sqrt{s-2(m_1^2+m_2^2)+(m_1^2-m_2^2)^2/s}}.
 \end{array}\nonumber
$
}
\end{eqnarray}


\begin{thebibliography}{99}

\bibitem{Dodelson:1993je} 
  S.~Dodelson and L.~M.~Widrow,
  Phys.\ Rev.\ Lett.\  {\bf 72}, 17 (1994).
  
  
 
\bibitem{Shi:1998km} 
  X.~D.~Shi and G.~M.~Fuller,
  Phys.\ Rev.\ Lett.\  {\bf 82}, 2832 (1999).



\bibitem{Boyarsky:2018tvu}
A.~Boyarsky, M.~Drewes, T.~Lasserre, S.~Mertens and O.~Ruchayskiy,
Prog. Part. Nucl. Phys. \textbf{104}, 1-45 (2019).



\bibitem{Lebedev:2023uzp}
O.~Lebedev and T.~Toma,
JHEP \textbf{05}, 108 (2023).


\bibitem{Asaka:2005an} 
  T.~Asaka, S.~Blanchet and M.~Shaposhnikov,
  Phys.\ Lett.\ B {\bf 631}, 151 (2005).
     
\bibitem{Asaka:2006nq} 
  T.~Asaka, M.~Laine and M.~Shaposhnikov,
  JHEP {\bf 0701}, 091 (2007);
  Erratum: [JHEP {\bf 1502}, 028 (2015)].


 \bibitem{Kusenko:2006rh} 
  A.~Kusenko,
  Phys.\ Rev.\ Lett.\  {\bf 97}, 241301 (2006).
  
\bibitem{DeRomeri:2020wng}
V.~De Romeri, D.~Karamitros, O.~Lebedev and T.~Toma,
JHEP \textbf{10}, 137 (2020).

 \bibitem{Coy:2022unt}
R.~Coy and M.~A.~Schmidt,
JCAP \textbf{08}, 070 (2022).





\bibitem{Hall:2009bx}
L.~J.~Hall, K.~Jedamzik, J.~March-Russell and S.~M.~West,
JHEP \textbf{03}, 080 (2010).


\bibitem{Lebedev:2022cic}
O.~Lebedev,
JCAP \textbf{02}, 032 (2023).


\bibitem{Koutroulis:2023fgp}
F.~Koutroulis, O.~Lebedev and S.~Pokorski,
``Gravitational production of sterile neutrinos,''
[arXiv:2310.15906 [hep-ph]], {\it to appear in JHEP}.



\bibitem{Cosme:2023xpa}
C.~Cosme, F.~Costa and O.~Lebedev,
``Freeze-in at stronger coupling,''
[arXiv:2306.13061 [hep-ph]], {\it to appear in Phys. Rev. D.}


 \bibitem{Hambye:2018dpi}
T.~Hambye, M.~H.~G.~Tytgat, J.~Vandecasteele and L.~Vanderheyden,
Phys. Rev. D \textbf{98}, no.7, 075017 (2018).


\bibitem{Giudice:2003jh}
G.~F.~Giudice, A.~Notari, M.~Raidal, A.~Riotto and A.~Strumia,
Nucl. Phys. B \textbf{685}, 89-149 (2004).


\bibitem{Falkowski:2015iwa}
A.~Falkowski, C.~Gross and O.~Lebedev,
JHEP \textbf{05}, 057 (2015).



\bibitem{Lebedev:2021xey}
O.~Lebedev,
Prog. Part. Nucl. Phys. \textbf{120}, 103881 (2021).



\bibitem{Hannestad:2004px}
S.~Hannestad,
Phys. Rev. D \textbf{70}, 043506 (2004).


\bibitem{Cosme:2024ndc}
C.~Cosme, F.~Costa and O.~Lebedev,
``Temperature evolution in the Early Universe and freeze-in at stronger coupling,''
[arXiv:2402.04743 [hep-ph]].



\bibitem{Belanger:2013ywg}
G.~Belanger, F.~Boudjema and A.~Pukhov,
``micrOMEGAs : a code for the calculation of Dark Matter properties in generic models of particle interaction,''
[arXiv:1402.0787 [hep-ph]].



\bibitem{Drewes:2015eoa}
M.~Drewes and J.~U.~Kang,
JHEP \textbf{05}, 051 (2016).

\bibitem{Konig:2016dzg}
J.~K\"onig, A.~Merle and M.~Totzauer,
JCAP \textbf{11}, 038 (2016).



\bibitem{Kolb:1990vq}
E.~W.~Kolb and M.~S.~Turner,
Front. Phys. \textbf{69}, 1-547 (1990).



\bibitem{Gondolo:1990dk}
P.~Gondolo and G.~Gelmini,
Nucl. Phys. B \textbf{360}, 145-179 (1991).


\bibitem{Bernstein:1985th}
J.~Bernstein, L.~S.~Brown and G.~Feinberg,
Phys. Rev. D \textbf{32}, 3261 (1985).


\bibitem{Silva-Malpartida:2023yks}
J.~Silva-Malpartida, N.~Bernal, J.~Jones-P\'erez and R.~A.~Lineros,
JCAP \textbf{09}, 015 (2023).



\bibitem{LZ:2022lsv}
J.~Aalbers \textit{et al.} [LZ],
Phys. Rev. Lett. \textbf{131}, no.4, 041002 (2023).


\bibitem{Kim:2006af}     
Y.~G.~Kim and K.~Y.~Lee,
Phys. Rev. D \textbf{75}, 115012 (2007).


\bibitem{Kim:2008pp}
Y.~G.~Kim, K.~Y.~Lee and S.~Shin,
JHEP \textbf{05}, 100 (2008).



\bibitem{Baek:2011aa}
S.~Baek, P.~Ko and W.~I.~Park,
JHEP \textbf{02}, 047 (2012).






\bibitem{Djouadi:2011aa}
A.~Djouadi, O.~Lebedev, Y.~Mambrini and J.~Quevillon,
Phys. Lett. B \textbf{709}, 65-69 (2012).



\bibitem{Lopez-Honorez:2012tov}
L.~Lopez-Honorez, T.~Schwetz and J.~Zupan,
Phys. Lett. B \textbf{716}, 179-185 (2012).



\bibitem{XENON:2020kmp}
E.~Aprile \textit{et al.} [XENON],
JCAP \textbf{11}, 031 (2020).
 
   
\bibitem{DARWIN:2016hyl}
J.~Aalbers \textit{et al.} [DARWIN],
JCAP \textbf{11}, 017 (2016).



\bibitem{ATLAS:2022eap}
G.~Aad \textit{et al.} [ATLAS],
JHEP \textbf{07}, 200 (2023).



\bibitem{Robens:2022cun}
T.~Robens,
Springer Proc. Phys. \textbf{292}, 141-152 (2023)
[arXiv:2209.15544 [hep-ph]].

\bibitem{Tania2}
Tania Robens, {\it private communication.}




\bibitem{Biekotter:2022ckj}
T.~Biek\"otter and M.~Pierre,
Eur. Phys. J. C \textbf{82}, no.11, 1026 (2022).


\bibitem{Bringmann:2021sth}
T.~Bringmann, S.~Heeba, F.~Kahlhoefer and K.~Vangsnes,
JHEP \textbf{02}, 110 (2022).



\bibitem{ATLAS:2020kdi}
 [ATLAS],
``Combination of searches for invisible Higgs boson decays with the ATLAS experiment,''
ATLAS-CONF-2020-052.



\bibitem{RivadeneiraBracho:2022sph}
P.~A.~Rivadeneira Bracho,
``Search for invisible decays of the Higgs boson produced via vector boson fusion at the ATLAS detector with 139 fb\ensuremath{-}1 of integrated luminosity,''
{\it PhD thesis, University of Hamburg, 2022.}


\bibitem{FCC}
P. Giacomelli, {\it talk at ICHEP 2018,} 
https://indico.cern.ch/event/686555/contributions/2971566/attachments/1682031/2703684/Higgs-measurements-FCC-ICHEP-2018\_169.pdf
 


\end{thebibliography}

\end{document}